\documentclass[journal]{IEEEtran}
\usepackage{times,amsmath,amsthm,epsfig,amssymb,graphicx,amsfonts}
\usepackage{subfigure}
\usepackage{psfrag}
\usepackage{ifpdf}
\usepackage{cite}
\usepackage{array}

\newtheoremstyle{mytheoremstyle} % name
    {\topsep}                    % Space above
    {\topsep}                    % Space below
    {}                           % Body font
    {}                           % Indent amount
    {\itshape}                   % Theorem head font
    {.}                          % Punctuation after theorem head
    {.5em}                       % Space after theorem head
    {}  % Theorem head spec (can be left empty, meaning ‘normal’)

\theoremstyle{mytheoremstyle}

\newtheorem{difinition}{Definition}

\newtheorem{lemma}{Lemma}
\newtheorem{theorem}{Theorem}
\newtheorem{corollary}{Corollary}

\title{Optimized Spline Interpolation}

\author{Ramtin~Madani,~\IEEEmembership{Student Member,~IEEE,} Ali~Ayremlou,~\IEEEmembership{Student Member,~IEEE,} Arash~Amini, Farrokh~Marvasti,~\IEEEmembership{Senior Member,~IEEE,}

\thanks{All authors are with Advanced Communications Research Institute (ACRI), the Department of Electrical Engineering, Sharif University of Technology, Tehran, Iran, e-mails: \{r\_madani , a\_ayremlou , arashsil\}@ee.sharif.edu, {marvasti@sharif.ir}}}% <-this % stops a space
\begin{document}
\maketitle

\begin{abstract}
In this paper, we investigate the problem of designing compact support interpolation kernels for a given class of signals. By using calculus of variations, we simplify the optimization problem from an infinite nonlinear problem to a finite dimensional linear case, and then find the optimum compact support function that best approximates a given filter in the least square sense ($\ell_2$ norm). The benefit of compact support interpolants is the low computational complexity in the interpolation process while the optimum compact support interpolant gaurantees the highest achivable Signal to Noise Ratio (SNR). Our simulation results confirm the superior performance of the proposed splines compared to other conventional compact support interpolants such as cubic spline.

%The goal of this paper is to design compact support basis spline functions that best approximate a given filter (e.g., an ideal Lowpass filter). The optimum function is found by minimizing the least square problem ($\ell_2$ norm of the difference between the desired and the approximated filters) by means of the calculus of variation; more precisely, the introduced splines give optimal filtering properties with respect to their time support interval. Both mathematical analysis and simulation results confirm the superiority of these splines.
\end{abstract}

% NOTE keywords are not used for conference papers so do not populate them

\section{Introduction}\label{introduction}
\IEEEPARstart{D}{ue} the existence of powerful digital tools, nowadays it is very common to convert the continuous time signals into the discrete form, and after processing the discrete signal, we can convert the discrete signal back to the original domain. The conversion of the continuous signal into the discerete domain is usually called the sampling process; the common form of sampling consists of taking samples directly from the cotinuous signal at equidistant time instants (uniform sampling). Although the samples are uniquely determined by the continuous function, there are infinite number of continuous signals which produce the same set of samples. The reconstrcution process is defined as selecting one of the infinite possibilities which satisfies certain constraints. For a given set of constraints, a proper sampling scheme is the one that establishes a one-to-one mapping between the discrete signals and the set of continuous fucntions that satisfy the constraints. One of the well-known constraints is the (finite support in fourier domain) condition \cite{slepian1976}. Due to the advances in the wavelet theory \cite{daubechies1988orthonormal, strang1996wavelets, mallat1999wavelet}, a new intrest in continuous-time modeling and spline interpolation has been generated. Multiresolution analysis \cite{Ville2005Mul, Mallat1989Mul}, self-similarity \cite{Flandrin1992Self, Unser2007Self}, and singularity analysis \cite{Mallat1992Singularity} are inseparable from a continuous-time interpolation. Although proper sampling schemes provide a one-to-one mapping between the continuous and discerete forms, we still need the tools to recover a continuous time signal from its samples. In fact, interpolation techniques using splines are one of the best options here.

%\IEEEPARstart{T}{he} conversion of continuous-time signals such as multimedia data with discrete and digitized samples is a common trend nowadays. This is mainly due to the existence of powerful tools in the discrete domain. However, the conversion of continuous-time signals into the discrete form by means of sampling  may destroy all or some parts of the data.
%Under certain conditions on the continuous signal, such as bandlimitedness \cite{slepian1976}, the sampling process is guaranteed to be one to one; i.e., there should be a priori a continuous model. In spite of the technological movement toward digital signal processing, by the advances in wavelet theory \cite{daubechies1988orthonormal, strang1996wavelets, mallat1999wavelet}, a revival of continuous-time modeling for the digital data has been triggered. Multiresolution analysis \cite{Ville2005Mul, Mallat1989Mul}, self-similarity \cite{Flandrin1992Self, Unser2007Self}, and singularity analysis \cite{Mallat1992Singularity} are inseparable from a continuous-time interpretation. It is therefore crucial to have efficient mathematical tools that allow easy switching from the digital domain to the continuous, and this is precisely the niche that splines, and, to some extent, wavelets, are trying to fill.

In this field, polynomial splines, such as B-Splines, are particularly popular; mainly due to their simplicity, compact support, and excellent 
approximation capabilities compared to other methods. B-Spline interpolations have spread to various applications \cite{ unser1999splines, unser1993b, unser2000sampling}. 

Many advantages of the B-splines arise from the fact that they are compact support functions. However, there is no evidence that they are the best compact support kernels for the interpolation process; i.e., it may be possible to improve the performance without compromising the desired property of the compact supportedness. In this paper, we focus on the problem of designing compact support splines that best resemple a given filter such as the ideal lowpass filter; more precisely, we aim to find a compact support spline that minimizes the least squared error when its cardinal spline is compared to a given fucntion. The given filter may be any arbitrary function that reflects the properties and constraints of the class of signals that enter the sampling process.

%Though B-Splines generate remarkable results in many applications, they are not the optimum solutions for filtering problems such as interpolation. This paper, focuses on the problem of designing optimal compact support splines which best approximate a given filter such as the ideal lowpass filter. In fact, the desired filter reflects the characteristics of the continuous-time model and can be arbitrary.

The remainder of the paper is organized as follows: The next section briefly describes the spline interpolation method. In section \ref{proposed}, a novel scheme is proposed to produce new optimized splines for interpolation regardless of the type of filtering. The performance of the proposed method is evaluated in section \ref{Simulation} by comparing the interpolation results of the proposed method to those of well-known interpolation techniques. Section \ref{conclusion} concludes the paper.

\section{Preliminaries}\label{Preliminaries}
\begin{figure}[t]
 \centering
 \includegraphics[width=90mm]{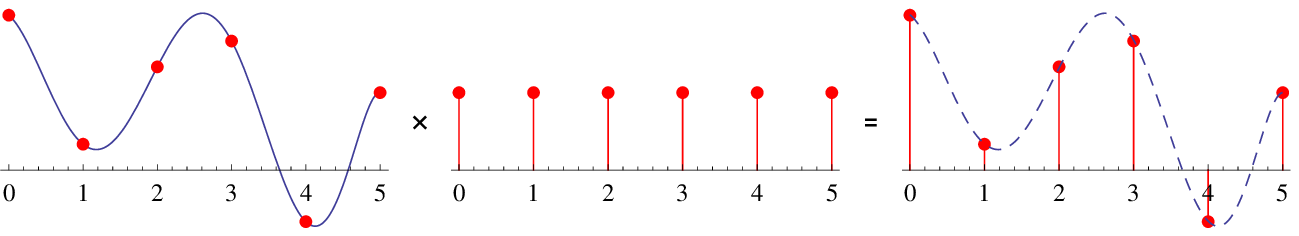}
 \caption{Sampling process modeled by multiplying an impulse train by a continuous time signal.}
 \label{fig1}
\end{figure}

In this paper, the following notations and definitions are used:
\begin{difinition}
For a continuous-time signal $x(t)$, a continuous-time signal $x_p(t)$ and a discrete-time signal $x_d[n]$ are defined as follows:
\begin{equation}
x_d[n]\triangleq x(nT)
\end{equation}
\begin{equation}
x_p(t)\triangleq x(t)p(t)=\sum_{n=-\infty}^{\infty}{x_d[n]\delta(t-nT)}
\end{equation}
where $p(t) \triangleq \sum_{n=-\infty}^{+\infty}{\delta(t-nT)}$ is the periodic impulse train.
\label{defps}
The sampling period $T$ is normalized to $1$ without any loss of generality.
\end{difinition}
The sampling process is shown in Fig \ref{fig1}.

\begin{difinition}
A Linear Time Invarient (LTI) filter with impulse response $h(t)$ is called to have interpolation property if and only if,
\begin{equation}
h_p(t)=\delta(t).
\end{equation}
\label{defIP}
\end{difinition} 
\begin{difinition}

For a continuous-time signal $x(t)$ and any odd integer $m$, $x^m_s(t)$ is a polynomial spline of degree $m$ if,
\begin{enumerate}
\item for any $n\in\mathbb{Z}$, $x^m_s(t)$ is a polynomial of (at most) degree $m$ in the interval $[n,n+1]$ (Compact support), 
\item for any $n\in\mathbb{Z}$, $x^m_s(n)=x_d[n]$ (Interpolation property),
\item $x^m_s\in C^{m-1}(-\infty,\infty)$ (Smoothness).
\end{enumerate}
\label{def2}
\end{difinition} 

According to the first property, $(m+1)^{th}$ derivative of $x^m_s$ is equal to an impulse train.
\begin{difinition}
For the polynomial spline $x^m_s(t)$, the polynomial spline coefficients $\dot{x}^m_d[n]$ are defined as:
\begin{equation}
\dot{x}^m_p(t)= \sum_ {n =-\infty}^{\infty}{\dot{x}^m_d[n]\delta(t-n)} \triangleq \frac{d^{m+1}}{{dt}^{m+1}}x^m_s(t)
\label{equ4}
\end{equation}
\label{coef}
\end{difinition}

\begin{figure}[t]
 \centering
 \includegraphics[width=90mm]{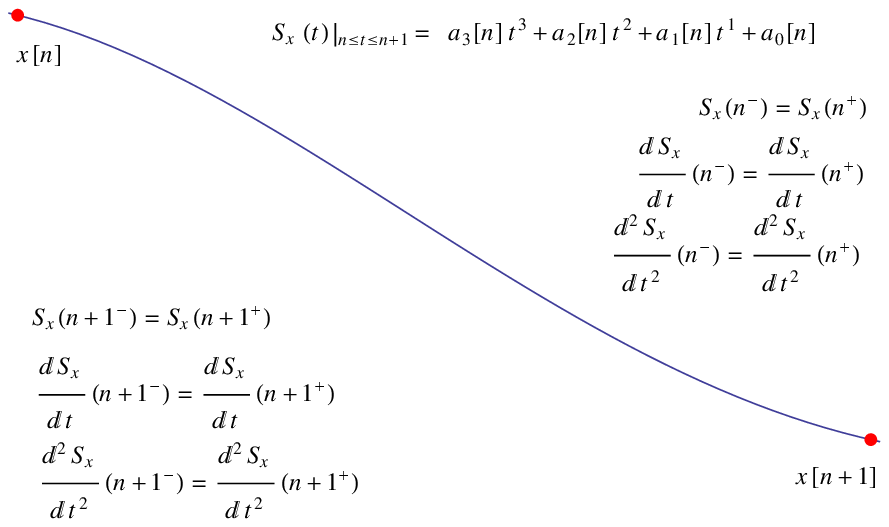}
 \caption{Spline of the degree $3$ conditions.}
 \label{fig2}
\end{figure}

\begin{figure}[t]
 \centering
 \includegraphics[width=90mm]{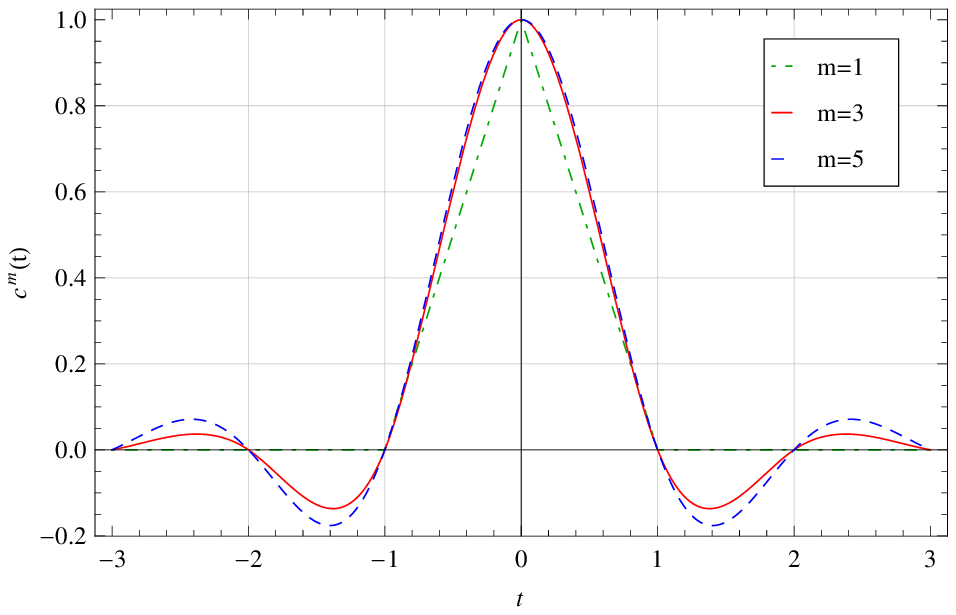}
 \caption{Cardinal splines of different degrees.}
 \label{fig3}
\end{figure}

To determine all polynomials of degree $m$ that form $x^m_s(t)$, the $m+1$ unknown spline coefficients should be found in order to satisfy the conditions 2 and 3 (Fig. \ref{fig2}). If the goal is to discover a piecewise polynomial signal that is $(m-1)$ times differentiable with continuous derivatives, a natural method is to derive $\dot{x}^m_d[n]$ according to $x_d[n]$ and then calculate the integral of $\dot{x}^m_x (t)$, $m+1$ times, i.e,
\begin{eqnarray}
x^m_s(t)&=&\int_{-\infty}^{t}{\int_{-\infty}^{t_{m}}{\dots\int_{-\infty}^{t_1}}}{\dot{x}^m_p(t_0){dt}_0\dots{dt}_{m-1}{dt}_m}\nonumber\\
&=&\left(u^{m+1}\ast \dot{x}^m_p\right)(t)
\label{Sx}
\end{eqnarray}
where $u^1(t)$ is the unity step function and for any $k\in\mathbb{N}$, $u^{k+1}(t)\triangleq \left(u^k\ast u^1\right)(t)$.
\begin{lemma}
If the Region Of Convergence (ROC) of $X^m_d(z)$ is not enclosed in the unit circle (i.e, there exists $z \in ROC \{X^m_d\}$ such that $\lvert z \rvert > 1$), then from Def. \ref{def2}, $x^m_s(t)$ will be uniquely determined from $x_d[n]$, and
\begin{equation}
x^m_s(t) = \left(\left(u^{m+1} \ast (u^{m+1}_p)^{-1} \right) \ast x_p \right) (t) \label{equ5}
\end{equation}
where $(u^{m+1}_p)^{-1}(t)$ is defined as the inverse of $u^{m+1}_p(t)$, i.e, $\left((u^{m+1}_p)^{-1} \ast u^{m+1}_p \right)(t)=\delta(t)$, and $X^m_d(z)$ is the z-transform of $x^m_d[n]$.\\
\begin{proof} From Def. \ref{coef},
\begin{eqnarray}
x^m_p(t)&=&x^m_s(t)p(t)\nonumber\\
&=&\left(u^{m+1}\ast \dot{x}^m_p\right)(t)p(t)\nonumber\\
&=&\left(u^{m+1}_p\ast \dot{x}^m_p\right)(t)
\end{eqnarray}
Hence,
\begin{equation}
x^m_d[n]=\left(u^{m+1}_d \ast \dot{x}^m_d\right)[n]
\end{equation}
The $ROC$ of $U^{m+1}_d(z)$ is $|z|>1$ and there are no zeros in this region. Since the $ROC$ of $X^m_d(z)$ in not enclosed in the unit circle, $(U^{m+1}_d)^{-1}(z)$ and $X^m_d(z)$ have a region in common. Thus,
\begin{equation}
\dot{x}^m_p(t)=\left((u^{m+1}_p)^{-1} \ast x^m_p\right)(t)
\end{equation}
and according to (\ref{Sx}),
\begin{eqnarray}
x^m_s(t) &=& \left(u^{m+1} \ast \dot{x}^m_p\right)(t) \nonumber\\ 
&=& \left(u^{m+1} \ast (u^{m+1}_p)^{-1}  \ast x^m_p \right) (t)
\end{eqnarray}
\end{proof}
\end{lemma}
\begin{difinition}
A discrete-time signal $y_d[n]$ is called a proper signal if and only if it is bounded and has a unique and bounded inverse $y_d^{-1}[n]$.
\end{difinition}
\begin{difinition}
For any continuous-time signal $y(t)$, if $y_d[n]$ is a proper signal, then $ \widehat{y}(t)$ is defined as follows:
\begin{equation}
\widehat{y}(t)=\left((y_p)^{-1}\ast y\right)(t)
\label{100}
\end{equation}
\label{def3}
\end{difinition}
\begin{corollary}
$\widehat{y}(t)$ in (\ref{100}) is the impulse response of a filter with interpolation property, in other words:
\begin{equation}
\widehat{y}_p(t)=\delta(t)
\end{equation}\\
\begin{proof}
\begin{eqnarray}
\widehat{y}_p(t)&=&\widehat{y}(t)p(t)\nonumber\\
&=&\left[\left((y_p)^{-1}\ast y\right)(t)\right]p(t)\nonumber\\
&=&\left((y_p)^{-1}\ast y_p\right)(t)= \delta(t)
\end{eqnarray}
\end{proof}
\end{corollary}
\begin{lemma}
For a given proper signal $y_d^m[n]$, the signal $\widehat{y_s^{m}}(t)$ is only a function of $m$ and does not depend on $y_d^m[n]$.
\\
\begin{proof}
\begin{eqnarray}
\widehat{y^m_s}(t)&=&\left(({(y^m_s)}_p)^{-1}\ast {y^m_s}\right)(t)\nonumber\\
&=&\left((y_p)^{-1}\ast {y^m_s}\right)(t)\nonumber\\
&=&\left((y_p)^{-1}\ast {\left(u^{m+1} \ast (u^{m+1}_p)^{-1} \right) \ast y_p}\right)(t)\nonumber\\
&=&{\left(u^{m+1} \ast (u^{m+1}_p)^{-1} \right) }(t) \label{equ13}
\end{eqnarray}
Hence $\widehat{y^m_s}(t)$ is only a function of $m$.
\end{proof}
\label{card}
\end{lemma}
\begin{difinition}
According to the above corollary, $c^m(t)\triangleq \widehat{y^m_s}(t)$ is defined as the cardinal spline of degree $m$ (Fig. \ref{fig3}).
\end{difinition}
From (\ref{equ5}) and (\ref{equ13}) it can be concluded that the polynomial spline interpolation is a linear shift invariant process according to $x_d[n]$ and its impulse response is represented by $c^m(t)$. $c^m(t)$, on the other hand, can be derived according to any arbitrary polynomial spline $y^m_s(t)$ such that $y^m_d(t)$ is a proper signal, i.e,
\begin{eqnarray}
{x^m_s}(t)&=&\left({c^m}\ast {x_p}\right)(t) \nonumber\\
&=&\left(y^m_s \ast \left((y_p)^{-1} \ast x_p \right)\right) (t)
\end{eqnarray}

The above equation divides the whole interpolation process into a discrete-time and a continuous-time section. If $y^m_s(t)$ is chosen as a time limited basis, both sections of this process can be extremely simplified and a considersble amount of continuous-time calculation can be avoided.
\begin{theorem}
$k=m+1$ is the least positive integer for which there exists a polynomial spline of degree $m$, such as $y^m_s(t)$, that vanishes outside the interval $(0,k)$, i.e,
\begin{eqnarray}
\forall{t} \notin (0,k) \Rightarrow y^m_s(t)=0
\label{equf}
\end{eqnarray}
then $k=m+1$.\\
\begin{proof}
Suppose that $y^m_s(t)$ satisfies (\ref{equf}) and ${\dot{Y}}^m_d(z)$ is the z-transform of ${\dot{y}^m_d[n]}$, where ${\dot{y}}^m_d[n]$  is the coefficients signal for the polynomial spline $y_s^m(t)$ according to the definition (\ref{coef}). It can be claimed that,
\begin{eqnarray}
(z-1)^{m+1}~|~{\dot{Y}}^m_d(z^{-1})
\label{aad}
\end{eqnarray}

In order to prove (\ref{aad}), we define the sequence of polynomials $\{Q_i\}^m_{i=0}$ such that $Q_0(z)\triangleq {\dot{Y}}^m_d(z^{-1})$ and for $1\leq n\leq m$, 
\begin{equation}
Q_i\triangleq z\frac{d}{dz}Q_{i-1}=\sum_{n=0}^{k}{\dot{y}^m_d[n] n^i z^n} 
\end{equation}
Also polynomial $H$ is defined as follows:
\begin{eqnarray}
H(t)&\triangleq& \frac{1}{m!} \sum_{i=0}^{m}{(-1)^i Q_i(1)\binom{m}{i}t^{m-i}}\nonumber\\
&=& \frac{1}{m!} \sum_{i=0}^{m}{(-1)^i\left(\sum_{n=0}^{k}{\dot{y}^m_d[n]n^i}\right)\binom{m}{i}t^{m-i}}\nonumber\\
&=& \frac{1}{m!} \sum_{n=0}^{k}{\dot{y}}^m_d[n]\sum_{i=0}^{m}\binom{m}{i}{(-1)^i {n^i} t^{m-i}}\nonumber\\
&=& \frac{1}{m!} \sum_{n=0}^{k}{{\dot{y}}^m_d[n](t-n)^m}
\label{H}
\end{eqnarray}
Thus, according to (\ref{Sx}), for any $t>k$, $H(t)$ is equal to $y^m_s(t)$, i.e,
\begin{equation}
\forall{t}>k \Rightarrow H(t)=y^m_s(t)=0
\end{equation}
Since $H$ is a polynomial that vanishes for infinite values of $t$, it should be equivalent to zero:
\begin{equation}
H(t)\equiv 0\Rightarrow Q_m(1)=Q_{m-1}(1)=\dots=Q_0(1)=0
\end{equation}
From the above equations, it is concluded directly by induction that,
\begin{eqnarray}
\forall{n \in \mathbb{N}}; 0\leq n\leq m \Rightarrow {\frac{d^n}{{dz}^n}{\dot{Y}}^m_d(z^{-1})}\Big|_{z=1}=0
\end{eqnarray}
Hence, $(z-1)^{m+1}$ divides ${\dot{Y}}^m_d(z^{-1})$.
On the other hand, since ${\dot{Y}}^m_d(z^{-1})=\sum_{n=0}^{k}{\dot{y}^m_d[n]z^n}$, (\ref{aad}) shows that $\dot{y}^m_d[m+1]\neq0$. Hence,
\begin{equation}
y^m_s(t) |_{t\in(m+1-\epsilon,m+1+\epsilon)}  \neq 0
\end{equation}
Thus $k \geq m+1$.

Finally, it must be shown that there exists a polynomial spline of degree $m$ that is supported in the interval $(0,m+1)$. Suppose that ${\dot{Y}}^m_d(z)=(z^{-1}-1)^{m+1}$, then
\begin{eqnarray}
&\Rightarrow& y^m_d[n]=(-1)^n\binom{m+1}{n} \\
&\Rightarrow& y^m_s(t)= u^{m+1} \ast \left[ \sum_{n=0}^{m+1} y^m_d[n] \delta(t-n) \right] \\
&\Rightarrow& y^m_s(t)= \sum_{n=0}^{m+1} y^m_d[n] u^{m+1}(t-n).
\end{eqnarray}
Thus, for all $t\geq m+1$, $y^m_s(t)=0$.
\end{proof}
\end{theorem}
\begin{difinition}
The polynomial B-Spline of degree $m$ is defined as follows:
\begin{equation}
\beta^m(t) \triangleq \sum_{n=0}^{m+1} (-1)^n\binom{m+1}{n} u^{m+1}(t-n)
\end{equation}
\label{BSpline}
\end{difinition}

Now according to the lemma \ref{card}, the polynomial spline interpolation process can be implemented by a compact support kernel such as the polynomial B-Splines, i.e,
\begin{eqnarray}
\dot{x}^m_d[n] = \left({(\beta^m_d)}^{-1} \ast x^m_d\right)[n] \label{inter1}\\
x^m_s(t) = \sum^{\infty}_{n=-\infty}{\dot{x}}^m_d[n]\beta^m{(t-n)} \label{inter2}
\end{eqnarray}

\section{Proposed Optimized B-Spline}\label{proposed}
In many applications, it is desirable that the interpolation filter resembles an ideal filter, and the second and third properties in Def. \ref{def2} may be ignored. In this section an optimized basis spline will be introduced to emulate a desired filter.
\begin{difinition} Let $D$ denote the set of all continuous-time signals that satisfy the dirichlet conditions, i.e, for any $y(t)\in D$
\begin{enumerate}
\item $y(t)$ has a finite number of extrema in any given interval,
\item $y(t)$ has a finite number of discontinuities in any given interval,
\item $y(t)$ is absolutely integrable over a period,
\item $y(t)$ is bounded.
\label{3-3}
\end{enumerate}
\end{difinition}

First of all, an affine subspace of all signals that satisfy the dirichlet conditions will be defined, and then the optimized solution will be obtained in this set by the calculus of variation.

\begin{difinition} Let $y_d[n]$ be a proper signal that vanishes for all $n\leq0$ and $n\geq m+1$, then $\chi^m(y_d)$ is the set of all continuous-time signals $y(t)$ that satisfy the following conditions,
\begin{enumerate}
\item $y \in D$
\item $\forall n \in \mathbb{N};~~ y(n)=y_d[n]$
\item $\forall t \notin (0,m+1);~~ y(t)=0$
\label{3-3}
\end{enumerate}
\end{difinition}

The use of $y(t) \in \chi^m(y_d)$ as a basis spline to interpolate $x_d[n]$ according to equations (\ref{inter1}) and (\ref{inter2}) is a linear time invariant process with the impulse response $\widehat{y}(t)$.

\begin{difinition} The error function $  e_x\colon \chi^m(y_d) \to \mathbb{R} $ is defined as follows:
\begin{eqnarray}
e_x(y) &\triangleq& \parallel \widehat{y} \ast x_p - x {\parallel}_2 \nonumber\\
&=& \int_{-\infty}^{\infty} {|    (\widehat{y} \ast x_p)(t) - x(t)    |^2 dt} \nonumber\\
&=& \int_{-\infty}^{\infty} {|   \mathcal{F}\{ \widehat{y} \ast x_p \}-\mathcal{F}\{ x \}   |^2 df} \nonumber\\
&=& \int_{-\infty}^{\infty} {|   \mathcal{F}\{ \left((y_p)^{-1}\ast y\right) \ast x_p \}-\mathcal{F}\{ x \}|^2 df}\nonumber\\
&=& \int_{-\infty}^{\infty} {| \frac{\mathcal{F}\{x_p\}}{\mathcal{F}\{y_p\}} \mathcal{F}\{y\}-\mathcal{F}\{x\}|^2 df}.
\end{eqnarray}
where $\mathcal{F}$ is defined as the continuous time Fourier transform operator.
\end{difinition}

\begin{difinition} According to the above definition, if $\rho^m_d$ is a proper signal that vanishes for all $n\leq0$ and $n\geq m+1$, an optimized basis spline $\rho^m [x,\rho^m_d]$ is defined as follows: 
\begin{equation}
\rho^m [x,\rho^m_d] \triangleq \underset{y\in \chi^m(\rho^m_d)}{\operatorname{arg\,min}} e_x(y) \label{op}
\end{equation}
\end{difinition}

Now, for a given proper discrete signal $\rho^m_d$ with the required vanishing property, we employ the calculus of variations in order to find the optimized continuous basis spline $\rho^m$ that minimizes the error function $e_x(\rho^m)$.

\begin{theorem}
Equation (\ref{op}) has a unique solution that satisfies the following property,
\begin{eqnarray}
[x_p*\overline{x_p}*({\rho}_p^m)^{-1}*(\overline{{\rho}_p^m})^{-1}]*{\rho}^m= [(\overline{\rho_p^m})^{-1}*\overline{x_p}]*x
\label{3-9}
\end{eqnarray}
for all $t\in(0,m+1)$, where $\overline{y}(t)\triangleq {y}^{\ast}(-t)$.\\
\begin{proof}
For $\gamma \in \chi^m(0)$ and any $\varepsilon>0$, we have $\rho^m+\varepsilon\gamma \in \chi^m(\rho^m_d)$, and the variational derivation of $e_x(\rho^m)$ with respect to $\rho^m$ with $\gamma$ as the test function is equal to
\begin{eqnarray}
\langle e_x(\rho^m),\gamma\rangle &\triangleq& \lim_{\varepsilon \rightarrow 0}\frac{e_x(\rho^m+\varepsilon\gamma)-e_x(\rho^m)}{\varepsilon}  \nonumber\\
&=&2\int\limits_{-\infty}^{\infty}\Re\{\gamma(t)\}\Re\Bigg\{\mathcal{F}^{-1}\Bigg\{ \left[\frac{\mathcal{F}\{x_p\}}{\mathcal{F}\{\rho_p^m\}}\right]^* \nonumber\\ 
&& \left[\frac{\mathcal{F}\{\rho^m\}}{\mathcal{F}\{\rho_p^m\}}\mathcal{F}\{x_p\}-\mathcal{F}\{x\}\right] \Bigg\}\Bigg\}dt \nonumber\\ 
&-&2\int\limits_{-\infty}^{\infty}\Im\{\gamma(t)\}\Im\Bigg\{\mathcal{F}^{-1}\Bigg\{ \left[\frac{\mathcal{F}\{x_p\}}{\mathcal{F}\{\rho_p^m\}}\right]^* \nonumber\\ 
&& \left[\frac{\mathcal{F}\{\rho^m\}}{\mathcal{F}\{\rho_p^m\}}\mathcal{F}\{x_p\}-\mathcal{F}\{x\}\right] \Bigg\}\Bigg\}dt
\label{3-6}
\end{eqnarray}

\begin{figure*}[t]
%\normalsize
\small
\begin{equation}
\label{eqn_dbl_x}
\begin{split}
\langle e_x(\rho^m[x,\rho^m_d]),\gamma\rangle
&\triangleq
\lim_{\varepsilon \rightarrow 0}\frac{e_x(\rho^m+\varepsilon\gamma)-e_x(\rho^m)}{\varepsilon}\\
&=
\lim_{\varepsilon \rightarrow 0}\frac{1}{\varepsilon}\left[\int_{-\infty}^{\infty}\left|\left(\frac{\mathcal{F}\{\rho^m+\varepsilon\gamma\}}{\mathcal{F}\{(\rho^m+\varepsilon\gamma)p\}}\right)\mathcal{F}\{x_p\}-\mathcal{F}\{x\}\right|^2df-\int_{-\infty}^{\infty}\left|\left(\frac{\mathcal{F}\{\rho^m\}}{\mathcal{F}\{\rho_p^m\}}\right)\mathcal{F}\{x_p\}-\mathcal{F}\{x\}\right|^2df\right]\\
&=
\lim_{\varepsilon \rightarrow 0}\frac{1}{\varepsilon} \int_{-\infty}^{\infty}\Bigg\{\left[\Re\left\{\left(\frac{\mathcal{F}\{\rho^m+\varepsilon\gamma\}}{\mathcal{F}\{(\rho^m+\varepsilon\gamma)p\}}\right)\mathcal{F}\{x_p\}-\mathcal{F}\{x\}\right\}^2-\Re\left\{\left(\frac{\mathcal{F}\{\rho^m\}}{\mathcal{F}\{\rho_p^m\}}\right)\mathcal{F}\{x_p\}-\mathcal{F}\{x\}\right\}^2\right]\\
&\quad
+\left[\Im\left\{\left(\frac{\mathcal{F}\{\rho^m+\varepsilon\gamma\}}{\mathcal{F}\{(\rho^m+\varepsilon\gamma)p\}}\right)\mathcal{F}\{x_p\}-\mathcal{F}\{x\}\right\}^2-\Im\left\{\left(\frac{\mathcal{F}\{\rho^m\}}{\mathcal{F}\{\rho_p^m\}}\right)\mathcal{F}\{x_p\}-\mathcal{F}\{x\}\right\}^2\right]\Bigg\}df\\
&=
\lim_{\varepsilon \rightarrow 0}\frac{1}{\varepsilon} \int_{-\infty}^{\infty}\Bigg\{\Re\left\{\left(\frac{\mathcal{F}\{\rho^m+\varepsilon\gamma\}}{\mathcal{F}\{(\rho^m+\varepsilon\gamma)p\}}-\frac{\mathcal{F}\{\rho^m\}}{\mathcal{F}\{\rho^m_p\}}\right)\mathcal{F}\{x_p\}\right\}\Re\left\{\left(\frac{\mathcal{F}\{\rho^m+\varepsilon\gamma\}}{\mathcal{F}\{(\rho^m+\varepsilon\gamma)p\}}+\frac{\mathcal{F}\{\rho^m\}}{\mathcal{F}\{\rho^m_p\}}\right)\mathcal{F}\{x_p\}-2\mathcal{F}\{x\}\right\}\\
&\quad
+\Im\left\{\left(\frac{\mathcal{F}\{\rho^m+\varepsilon\gamma\}}{\mathcal{F}\{(\rho^m+\varepsilon\gamma)p\}}-\frac{\mathcal{F}\{\rho^m\}}{\mathcal{F}\{\rho^m_p\}}\right)\mathcal{F}\{x_p\}\right\}\Im\left\{\left(\frac{\mathcal{F}\{\rho^m+\varepsilon\gamma\}}{\mathcal{F}\{(\rho^m+\varepsilon\gamma)p\}}+\frac{\mathcal{F}\{\rho^m\}}{\mathcal{F}\{\rho^m_p\}}\right)\mathcal{F}\{x_p\}-2\mathcal{F}\{x\}\right\}\Bigg\}df\\
&=
\lim_{\varepsilon \rightarrow 0}\frac{1}{\varepsilon} \int_{-\infty}^{\infty}\Bigg\{\Re\left\{\left(\frac{\mathcal{F}\{\rho^m+\varepsilon\gamma\}-\mathcal{F}\{\rho^m\}}{\mathcal{F}\{\rho^m_p\}}\right)\mathcal{F}\{x_p\}\right\}\Re\left\{\left(\frac{\mathcal{F}\{\rho^m+\varepsilon\gamma\}+\mathcal{F}\{\rho^m\}}{\mathcal{F}\{\rho^m_p\}}\right)\mathcal{F}\{x_p\}-2\mathcal{F}\{x\}\right\}\\
&\quad
+\Im\left\{\left(\frac{\mathcal{F}\{\rho^m+\varepsilon\gamma\}-\mathcal{F}\{\rho^m\}}{\mathcal{F}\{\rho^m_p\}}\right)\mathcal{F}\{x_p\}\right\}\Im\left\{\left(\frac{\mathcal{F}\{\rho^m+\varepsilon\gamma\}+\mathcal{F}\{\rho^m\}}{\mathcal{F}\{\rho^m_p\}}\right)\mathcal{F}\{x_p\}-2\mathcal{F}\{x\}\right\}\Bigg\}df\\
&=
\lim_{\varepsilon \rightarrow 0}\frac{1}{\varepsilon} \int_{-\infty}^{\infty}\Bigg\{\Re\left\{\left(\frac{\mathcal{F}\{\varepsilon\gamma\}}{\mathcal{F}\{\rho^m_p\}}\right)\mathcal{F}\{x_p\}\right\}\Re\left\{\left(\frac{\mathcal{F}\{2\rho^m+\varepsilon\gamma\}}{\mathcal{F}\{\rho^m_p\}}\right)\mathcal{F}\{x_p\}-2\mathcal{F}\{x\}\right\}\\
&\quad
+\Im\left\{\left(\frac{\mathcal{F}\{\varepsilon\gamma\}}{\mathcal{F}\{\rho^m_p\}}\right)\mathcal{F}\{x_p\}\right\}\Im\left\{\left(\frac{\mathcal{F}\{2\rho^m+\varepsilon\gamma\}}{\mathcal{F}\{\rho^m_p\}}\right)\mathcal{F}\{x_p\}-2\mathcal{F}\{x\}\right\}\Bigg\}df\\
&=
\int\limits_{-\infty}^{\infty}\Re\left\{\frac{\mathcal{F}\{\gamma\}}{\mathcal{F}\{\rho^m_p\}}\mathcal{F}\{x_p\}\right\}\Re\left\{2\frac{\mathcal{F}\{\rho^m\}}{\mathcal{F}\{\rho^m_p\}}\mathcal{F}\{x_p\}-2\mathcal{F}\{x\}\right\}+\Im\left\{\frac{\mathcal{F}\{\gamma\}}{\mathcal{F}\{\rho^m_p\}}\mathcal{F}\{x_p\}\right\}\Im\left\{2\frac{\mathcal{F}\{\rho^m\}}{\mathcal{F}\{\rho^m_p\}}\mathcal{F}\{x_p\}-2\mathcal{F}\{x\}\right\}df\\
&=
2\Re\left\{\int_{-\infty}^{\infty}\left[\frac{\mathcal{F}\{\gamma\}}{\mathcal{F}\{\rho^m_p\}}\mathcal{F}\{x_p\}\right]\left[\frac{\mathcal{F}\{\rho^m\}}{\mathcal{F}\{\rho^m_p\}}\mathcal{F}\{x_p\}-\mathcal{F}\{x\}\right]^*df\right\}\\
&=
2\Re\left\{\int_{-\infty}^{\infty}\mathcal{F}\{\gamma\}\left[\frac{\mathcal{F}\{x_p\}}{\mathcal{F}\{\rho^m_p\}}\right]\left[\frac{\mathcal{F}\{\rho^m\}}{\mathcal{F}\{\rho^m_p\}}\mathcal{F}\{x_p\}-\mathcal{F}\{x\}\right]^*df\right\}\\
&=
2\Re\Bigg\{\int_{-\infty}^{\infty}\gamma(t)\mathcal{F}^{-1}\left\{\left[\frac{\mathcal{F}\{x_p\}}{\mathcal{F}\{\rho^m_p\}}\right]^*\left[\frac{\mathcal{F}\{x_p\}}{\mathcal{F}\{\rho^m_p\}}\mathcal{F}\{\rho^m\}-\mathcal{F}\{x\}\right]\right\}\Bigg\}dt\\
&=
2\int_{-\infty}^{\infty}\Re\{\gamma(t)\}\Re\Bigg\{\mathcal{F}^{-1}\left\{\left[\frac{\mathcal{F}\{x_p\}}{\mathcal{F}\{\rho^m_p\}}\right]^*\left[\frac{\mathcal{F}\{x_p\}}{\mathcal{F}\{\rho^m_p\}}\mathcal{F}\{\rho^m\}-\mathcal{F}\{x\}\right]\right\}\Bigg\}dt\\
&\quad
-2\int_{-\infty}^{\infty}\Im\{\gamma(t)\}\Im\Bigg\{\mathcal{F}^{-1}\left\{\left[\frac{\mathcal{F}\{x_p\}}{\mathcal{F}\{\rho^m_p\}}\right]^*\left[\frac{\mathcal{F}\{x_p\}}{\mathcal{F}\{\rho^m_p\}}\mathcal{F}\{\rho^m\}-\mathcal{F}\{x\}\right]\right\}\Bigg\}dt\\
\end{split}
\end{equation}
\hrulefill
\vspace*{4pt}
\end{figure*}

The proof of (\ref{3-6}) is presented in (34). Since $\chi^m(\rho^m_d)$ is boundless, in order to minimize $e_x(\rho^m)$, $ \langle e(\rho^m),\gamma\rangle$ should be zero for all $\gamma \in \chi^m(0)$, which implies that the second term inside the integrals should be zero for $t\in(0,m+1)$, i.e,

\begin{equation}
\mathcal{F}^{-1}\Bigg\{ \left[\frac{\mathcal{F}\{x_p\}}{\mathcal{F}\{\rho_p^m\}}\right]^* \left[\frac{\mathcal{F}\{\rho^m\}}{\mathcal{F}\{\rho_p^m\}}\mathcal{F}\{x_p\}-\mathcal{F}\{x\}\right] \Bigg\}=0
\end{equation}
The above equation directly yields (\ref{3-9}).
\end{proof}
\end{theorem}
Thus, it is proven that the optimized basis spline which gives the best estimation of $x$, should satisfy (\ref{3-9}). By defining 
\begin{eqnarray}
v_p(t)&\triangleq&(x_p*\overline{x_p})*[(\rho^m_p)^{-1}*(\overline{\rho^m_p})^{-1}]\label{v}\\
w(t)&\triangleq&[(\overline{\rho^m_p})^{-1}*\overline{x_p}]*x,
\end{eqnarray}
we can rewrite (\ref{3-9}) as $(v_p*\rho^m)(t)=w(t)|_{t\in(0,m+1)}$. Since this equation is only valid in a particular interval, $(v_p)^{-1}$ cannot be used to obtain $\rho^m$. 
However, since $v_p$ is an impulse train (convolution of four impulse trains) 
%\textcolor{red}{which has the Hermitian symmetry according to (\ref{v})  (i.e. $v_p(t)=\overline{v_p}(t)$), 
we will show that the continuous functional equation in (\ref{3-9}) boils down to solving a finite Hermitian system of linear equations.
%which could be represented as a matrix form.
For this purpose, we first define the two following sequences of functions which are supported only on a unit-length interval: 

\begin{eqnarray}
R_n(t)&=&
\begin{cases}
\rho^m(t+n) &0\leq t< 1\\
0 &\text{o.w.}
\end{cases}\label{3-10}\\
W_n(t)&=&
\begin{cases}
w(t+n) &0 \leq t < 1\\
0 &\text{o.w.}
\end{cases}\label{3-11}
\end{eqnarray}
Now, here is (\ref{3-9}) in the matrix form: 
\begin{equation}
\left[\begin{smallmatrix}
\ v_d[0] & v_d[-1] & \dots & v_d[-m]\\
\ v_d[1] & v_d[0] & \dots & v_d[-m+1]\\
\vdots & \vdots & \ddots & \vdots\\
\ v_d[m] & v_d[m-1]  & \dots & v_d[0]
\end{smallmatrix}\right]
\left[\begin{smallmatrix}
R_{0}\\
R_{1}\\
\vdots\\
R_{m}
\end{smallmatrix}\right]=\left[\begin{smallmatrix}
W_{0}\\
W_{1}\\
\vdots\\
W_{m}
\end{smallmatrix}\right]\label{3-12}
\end{equation}
Since $v_d[-n]=v_d^*[n]$, the above matrix $V\triangleq [v_d[i-j]]_{i=1,\dots,m+1;j=1,\dots,m+1}$ is Hermitian also. Now by solving (\ref{3-12}), ${\{R_n\}}_{n=0}^m$ is derived and thus the optimized basis spline is given as
\begin{equation}
\\ \rho^m(t)=\sum_{n=0}^{m}R_n(t-n)
\label{3-14}
\end{equation}

%\begin{figure}[t]
% \centering
% \includegraphics[width=90mm]{Figures/Figure04.eps}
% \caption{The optimized spline versus cubic B-spline. $\rho^3\{h,\rho^3_d\}$ is the optimized basis spline built for estimating the ideal lowpass filter $h(t)=\frac{\sin(\pi t)}{\pi t}$ with $\rho^3_d(z)=0.235z+0.484z^2+0.235z^3$}.
% \vspace{-.2in}
% \label{fig:BsplineShapes}
% 
%\end{figure}
%\begin{figure}[t]
% \centering
% \includegraphics[width=90mm]{Figures/Figure05.eps}
% \caption{The comparison of the proposed method and the cubic spline for ideal lowpass filter.}
% \label{fig:InterpSplines}
%\end{figure}

This optimized basis spline minimizes the mean squared error of interpolation. On the other hand, smoothness of the optimized basis spline and causality of the prefilter can be achieved by adjusting $\rho_d$. 

Another application of (\ref{3-9}) is to approximate an ideal interpolation filter by an optimized basis spline. In fact, basis splines are superior to FIR filters.   

Now the goal is to design $\rho^m$ such that $\widehat{\rho^m}$ would be the best estimation of $h$, which denotes the impulse response of  a filter that has the interpolation property.

\begin{lemma}
(Estimating a desired filter) Assume $h(t)$ is the impulse response of a linear time invariant filter which satisfies the interpolation property and let $\rho_d^m[n]$ be a proper discrete signal, then,
\begin{equation}
\underset{y\in \chi^m(\rho^m_d)}{\operatorname{arg\,min}} {\lVert h-\widehat{y} \rVert}_2={\rho}^m [h,\rho^m_d] \label{op2}
\end{equation}

\begin{proof}
\begin{eqnarray}
e_h(y) &=& \int_{-\infty}^{\infty} {|   \mathcal{F}\{ \widehat{y} \ast h_p \}-\mathcal{F}\{ h \}   |^2 df} \nonumber\\
&=& \int_{-\infty}^{\infty} {| \mathcal{F}\{ \widehat{y}  \} \mathcal{F}\{h_p\}-\mathcal{F}\{h\}|^2 df} \nonumber\\
&=& \int_{-\infty}^{\infty} {| \mathcal{F}\{ \widehat{y}  \} -\mathcal{F}\{h\}|^2 df} \nonumber\\
&=& {\lVert h-\widehat{y} \rVert}_2
\end{eqnarray}
Thus
\begin{eqnarray}
\underset{y\in \chi^m(\rho^m_d)}{\operatorname{arg\,min}} {\lVert h-\widehat{y} \rVert}_2=\underset{y\in \chi^m(\rho^m_d)}{\operatorname{arg\,min}} e_h(y)={\rho}^m [h,\rho^m_d]
 \label{op3}
\end{eqnarray}
\end{proof}
\end{lemma}

%\begin{figure*}[t]
%\centering
%\subfigure[]
%{\label{fig:Original}\includegraphics[width=50mm]{Figures/Original.eps}}
%\subfigure[]
%{\label{fig:Bilinear}\includegraphics[width=50mm]{Figures/Lin.eps}}
%\subfigure[]
%{\label{fig:Bicubic}\includegraphics[width=50mm]{Figures/Cub.eps}}
%\subfigure[]
%{\label{fig:WZPCS}\includegraphics[width=50mm]{Figures/WZPCS.eps}}
%\subfigure[]
%{\label{fig:SAI}\includegraphics[width=50mm]{Figures/SAI.eps}} 
%\subfigure[]
%{\label{fig:Opt}\includegraphics[width=50mm]{Figures/Opt.eps}}
%\caption{Comparison of different methods for the Lena image: (a) The original image, (b) bilinear interpolation, (c) bicubic Interpolation. (d) WZP Cycle-Spinning \cite{temizel2005wavelet}, (e) SAI \cite{zhang2008image}, and (f) the proposed method.
%}
%\vspace{-.1in}
%\label{fig:Lena}
%\end{figure*}

\begin{corollary}
For an impulse response $h(t)$ with the interpolation property, if $\rho^m\in \chi^m(\rho_d^m)$ denotes the optimum basis spline for which $\widehat{\rho^m}$ best approximates $h(t)$ (i.e., $\widehat{\rho^{m}}(t)=\underset{y\in \chi^m(\rho^m_d)}{\operatorname{arg\,min}} {\lVert h-\widehat{y} \rVert}_2$), then $\rho^m\in \chi^m(\rho_d^m)$ satisfies:
\begin{eqnarray}
[({\rho}_p^m)^{-1}*(\overline{{\rho}_p^m})^{-1}]*{\rho}^m= [(\overline{\rho_p^m})^{-1}]*h
\label{3-19}
\end{eqnarray}
The proof follows directly from (\ref{3-9}) and the fact that $h(t)$ has the interpolation property.
\end{corollary}

Figure \ref{fig:BsplineShapes} shows the cubic B-spline and the optimized B-spline designed for estimating the ideal lowpass filter $h(t)=sinc(t)$ with $\rho^3_d(z)=0.235z+0.484z^2+0.235z^3$, while Fig. \ref{fig:InterpSplines} shows $\widehat{\rho^3}[h,\rho^3_d](t)$ in comparison to $c^3(t)$.

\section{Simulation Results}\label{Simulation}

\begin{figure}[t]
 \centering
 \includegraphics[width=90mm]{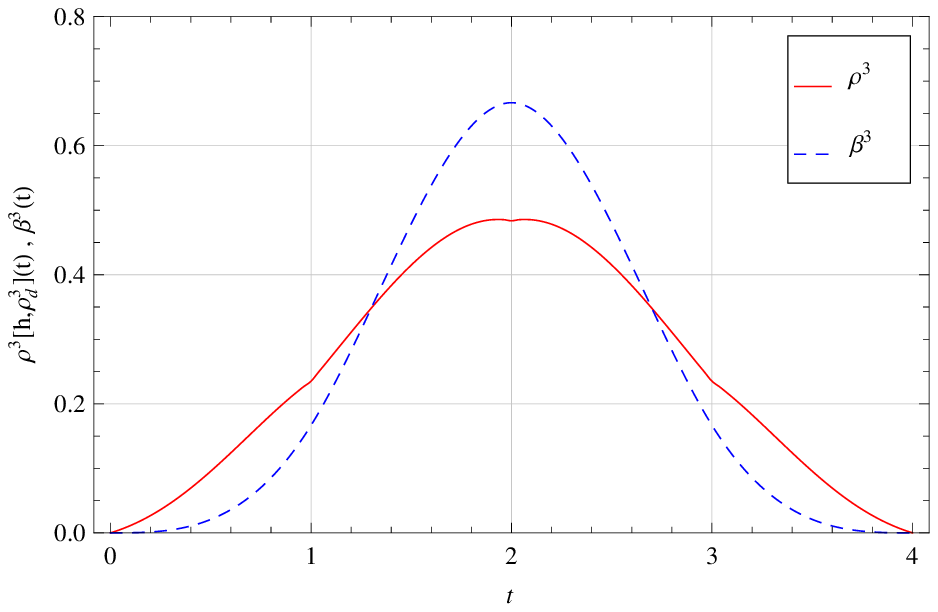}
 \caption{The optimized spline versus cubic B-spline. $\rho^3\{h,\rho^3_d\}$ is the optimized basis spline built for estimating the ideal lowpass filter $h(t)=\frac{\sin(\pi t)}{\pi t}$ with $\rho^3_d(z)=0.235z+0.484z^2+0.235z^3$.}
 \vspace{-.2in}
 \label{fig:BsplineShapes}
 
\end{figure}
\begin{figure}[t]
 \centering
 \includegraphics[width=90mm]{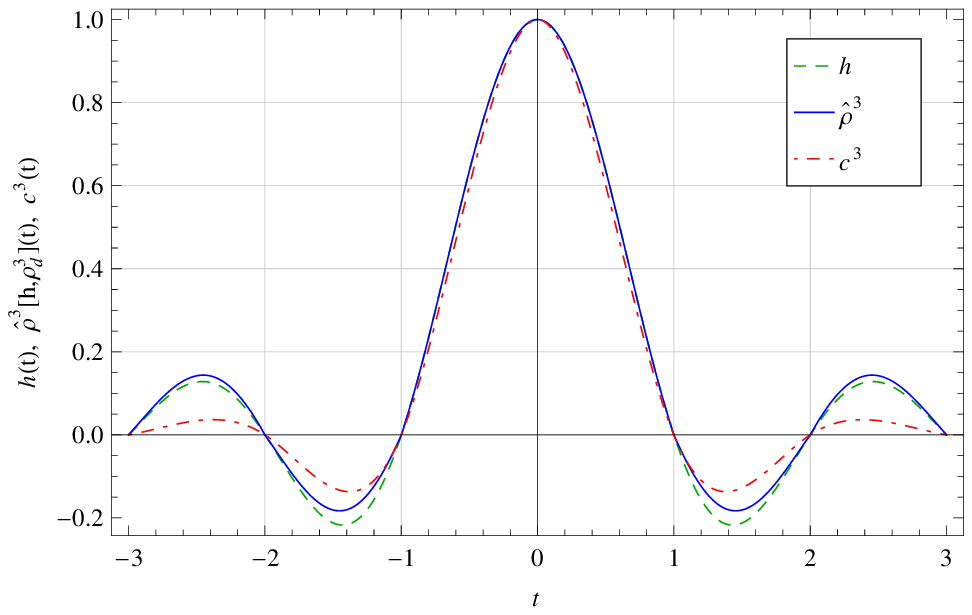}
 \caption{The comparison of the proposed method with the cubic spline for an ideal lowpass filter.}
 \label{fig:InterpSplines}
\end{figure}

To compare the proposed method with the existing interpolation techniques, we have performed various simulations. The cubic B-spline, due to its short time support and relatively high accuracy in approximating the ideal lowpass filter, is the most common technique for interpolating 1-D lowpass signals. For the purpose of comparison, we have optimized a spline function with the same time support as an ideal lowpass filter. Figures \ref{fig:BsplineShapes} and \ref{fig:InterpSplines} show the shape of the obtained B-spline and the interpolating spline, respectively. Figure \ref{fig:BsplineShapes} shows that the energy is more concentrated in the middle of the cubic B-spline while the optimized spline has slower decaying rate of energy at the sides. The resultant interpolation splines, as depicted in Fig. \ref{fig:InterpSplines}, reveal that the main advantage of the optimized spline ($\hat{\rho}^3$) compared to the cubic spline ($c^3$), is the smaller error in the first side-lobe. The SNR values of $\hat{\rho}^3$ and $c^3$ with respect to the $sinc$ function are $20.39$ and $13.15$dBs, respectively.

%The performance of the optimized Spline for an ideal lowpass filter has been compared to the B-spline and the results are depicted in Figs. \ref{fig:figure1} and \ref{fig6}. Fig. \ref{fig5} shows both optimized basis spline ${\rho}^3 [sinc(t),\rho^3_d]$ and the cubic B-spline. Fig. \ref{fig6} shows $\widehat{\rho}^3 [sinc(t),\rho^3_d]$ as compared to $c^3(t)$. The optimized spline is superior to the B-spline method since comparing $sinc(t)$ to $\widehat{\rho}^3 [sinc(t),\rho^3_d]$ and $c^3(t)$ the SNR values of these methods are $20.39$dB and $13.15$dB respectively.

\begin{table*}[t]
\begin{center}
\caption{PSNR (dB) Results of the Reconstructed Images by Various Methods, The Original Image was Anti-Aliased Before Sampling and The Results are Compared to The \textbf{Original} Image (Image Enlargement from $256\times256$ to $512\times512$)\label{T1}}
\begin{tabular}{|c|c|c|c|c|>{\centering}m{3cm}|m{4cm}<{\centering}|}
\hline
\textbf{Images} &Bilinear&Bicubic\cite{keys1981cubic}&WZP--CS\cite{temizel2005wavelet}&SAI \cite{zhang2008image}&Optimized Spline for Image ${\rho}^3 [x,\rho^3_d]$ &Optimized Spline for the Ideal LowPass ${\rho}^3 [sinc(t),\rho^3_d]$ \\\hline\hline
\textbf{Lena} & 30.33 & 30.44 & 30.12 & 30.96 & \textbf{32.46} & 32.20\\\hline
\textbf{Baboon} & 22.40 & 22.52 & 22.41 & 22.89 & 22.20 & \textbf{24.22}\\\hline
\textbf{Barbara} & 24.39 & 24.42 & 24.34 & 24.65 & 24.33 &  \textbf{25.14}\\\hline
\textbf{Peppers} & 29.46 & 29.46 & 29.14 & 29.72  & \textbf{31.14} & 31.04\\\hline
\textbf{Girl} & 30.98 & 30.98 & 30.66 & 30.43 & \textbf{31.70} & 30.87\\\hline
\textbf{Fishing bout} & 27.68 & 27.48 & 28.12 & 30.92 & 28.61 & \textbf{29.71}\\\hline
\textbf{Couple} & 27.35 & 27.50 & 27.27 & 27.39 & 28.04 & \textbf{28.97}\\\hline\hline
\textbf{Overall Average} & 27.51 & 27.54 & 27.43 & 28.14 & 28.35 & \textbf{28.88}\\\hline
\end{tabular}
\end{center}
\end{table*}

\begin{table*}[t]
\begin{center}
\caption{PSNR (dB) Results of the Reconstructed Images by Various Methods, The Original Image was Anti-Aliased Before Sampling and The Results are Compared to The \textbf{Anti-Aliased} Image (Image Enlargement from $256\times256$ to $512\times512$)\label{T2}}
\begin{tabular}{|c|c|c|c|c|>{\centering}m{3cm}|m{4cm}<{\centering}|}
\hline
\textbf{Images} &Bilinear&Bicubic\cite{keys1981cubic}&WZP--CS\cite{temizel2005wavelet}&SAI \cite{zhang2008image}&Optimized Spline for Image ${\rho}^3 [x,\rho^3_d]$ & Optimized Spline for the Ideal LowPass ${\rho}^3 [sinc(t),\rho^3_d]$ \\\hline\hline
\textbf{Lena} & 31.56 & 31.72 & 31.62 & 32.45 & \textbf{35.28} & 35.21 \\\hline
\textbf{Baboon} & 26.88 & 27.26 & 26.89 & 28.48 & 26.33 & \textbf{35.70} \\\hline
\textbf{Barbara} & 30.61 & 30.74 & 30.65 & 31.86 & 30.40 &  \textbf{35.72} \\\hline
\textbf{Peppers} & 31.62 & 31.82 & 31.77 & 32.14 & 37.41 & \textbf{38.02} \\\hline
\textbf{Girl} & 34.08 & 34.24 & 34.25 & 33.10 & \textbf{35.89} & 34.28 \\\hline
\textbf{Fishing bout} & 29.91 & 30.19 & 29.93 & 30.92 & 32.00 & \textbf{34.87} \\\hline
\textbf{Couple} & 29.81 & 30.10 & 29.84 & 29.77 & 31.44 & \textbf{34.28} \\\hline\hline
\textbf{Overall Average} & 30.64 & 30.87 & 30.71 & 31.25 & 32.68 & \textbf{35.44}\\\hline
\end{tabular}
\end{center}
\end{table*}

\begin{table*}[t]
\begin{center}
\caption{PSNR (dB) Results of the Reconstructed Images by Various Methods, The \textbf{Original Image was not Anti-Aliased} and The Results Are Compared to The \textbf{Original} Image $256\times256$ to $512\times512$)\label{T3}}
\begin{tabular}{|c|c|c|c|c|>{\centering}m{3cm}|m{4cm}<{\centering}|}
\hline
\textbf{Images} &Bilinear&Bicubic\cite{keys1981cubic}&WZP--CS\cite{temizel2005wavelet}&SAI \cite{zhang2008image}&Optimized Spline for Image ${\rho}^3 [x,\rho^3_d]$ &Optimized Spline for the Ideal LowPass ${\rho}^3 [sinc(t),\rho^3_d]$\\\hline\hline
\textbf{Lena} & 30.21 & 30.13 & 30.05 & 30.88 & \textbf{32.29} & 30.95\\\hline
\textbf{Baboon} & 21.67 & 21.34 & 21.70 & 22.09 & \textbf{22.50} & 21.63\\\hline
\textbf{Barbara} & 23.90 & 23.32 & 23.88 & 23.71 &  \textbf{25.10} & 22.58\\\hline
\textbf{Peppers} & 28.82 & 28.61 & 26.93 & 28.91 &  \textbf{30.64} & 29.77\\\hline
\textbf{Girl} & 30.41 & 29.97 & 30.20 & 29.94 & \textbf{30.90} & 29.20\\\hline
\textbf{Fishing bout} & 27.10 & 26.93 & 27.07 & 27.63 & \textbf{28.50} & 27.66\\\hline
\textbf{Couple} & 26.92 & 26.73 & 26.86 & 26.93 & \textbf{27.91} & 27.08\\\hline\hline
\textbf{Overall Average} & 27.00 & 26.72 & 26.67 & 27.16 & \textbf{29.12} & 26.98\\\hline
\end{tabular}
\end{center}
\end{table*}

For a more realistic comparison, we have applied different interpolation techniques on standard test images. For this purpose, the original images, with or without applying the anti-aliasing filter (ideal lowpass filter), are down-sampled by a factor 2 in each direction ($25\%$ of the original pixels) and then they are enlarged (zooming) using the interpolation techniques. The comparison is made with the following interpolation methods: 1) bilinear interpolation, 2) bicubic interpolation, 3) wavelet-domain zero padding cycle-spinning \cite{temizel2005wavelet}, and 4) soft-decision estimation technique for adaptive image interpolation \cite{zhang2008image}. Also for our proposed method, two different scenarios are implemented: the ideal filter for which we are optimizing the spline function is first taken as a $sinc$ filter, and first the spectrum of the original image.
%To consider practical applications, the method was tested on several standard monochrome images. These images are downsampled to provide the low solution images for interpolation. 
%In image applications, splines can be used for zooming and enlargements. 
%For comparison, four other image interpolation methods are also simulated: 1-bilinear interpolation, 2-bicubic interpolation, 3-wavelet-domain zero padding cycle-spinning \cite{temizel2005wavelet} and 4-soft-decision estimation technique for adaptive image interpolation \cite{zhang2008image}.
In order to evaluate the quality of the interpolated images, we have considered the Peak Signal-to-Noise Ratio (PSNR) criterion. Table \ref{T1} indicates the resultant PSNR values when the original image is subject to the anti-aliasing filter before down-sampling while the error is calculated based on the image without applying the filter. Table \ref{T2} contains similar values while the basis for the error calculation is the anti-aliased image.
In both cases, the PSNR criterion favors the proposed optimized spline for the ideal lowpass ($sinc$) filter. On the average, the proposed method out performs the other standard interpolating methods by $0.74$dB in Table \ref{T1} and $4.19$dB in Table \ref{T2}.

%Tables \ref{T1}, \ref{T2} and \ref{T3} show the Peak Signal-to-Noise Ratio (PSNR) performance of these four methods when applied to the seven well-known test images. In tables \ref{T1} and \ref{T2} the original image was anti-aliased before sampling and the results are compared to the original image and the anti-aliased image respectively.
%In both cases the proposed optimized spline for the ideal lowpass performes best among all methods. The proposed method performs in average over $1$dB and $4$dB improvement in table \ref{T1} and \ref{T2} respectively.

To exclude the effect of the anti-aliasing filter, the simulations are repeated without applying it and the results are presented in Table \ref{T3}. As expected, the optimized spline which is matched to the spectrum of the original image outperforms other competitors in all cases. It should be mentioned that the function ${\rho}^m [x,\rho^m_d]$ is not a universal filter in this case and depends on the choice of the image.

%Table \ref{T1} is also shows the PSNR performance when the original image was not anti-aliased before sampling and the results are compared to the original image.
%According to the definition of ${\rho}^m [x,\rho^m_d]$, in all cases the optimized spline for signal performes best among all methods, however the optimized spline for ideal lowpas is not necessarily the best.

\begin{figure*}[t]
\centering
\subfigure[]
{\label{fig:Original}\includegraphics[width=50mm]{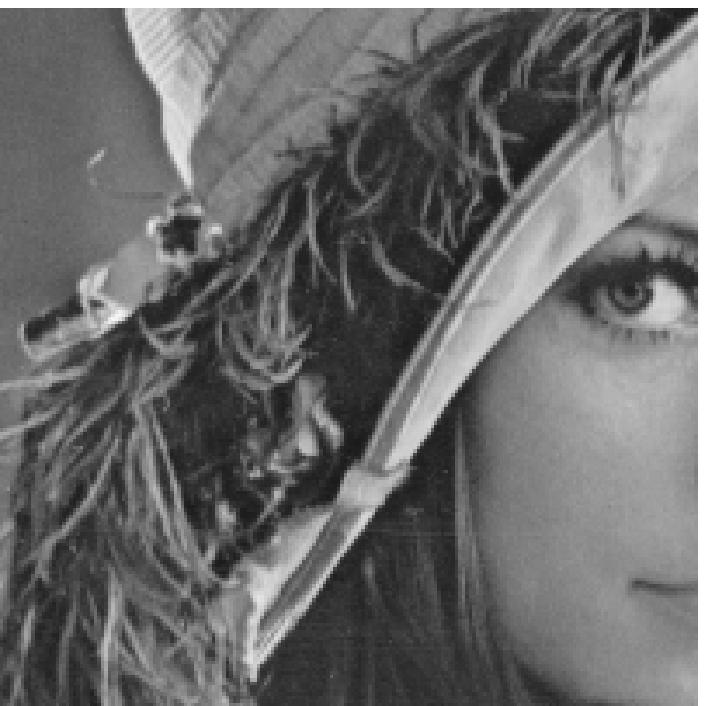}}
\subfigure[]
{\label{fig:Bilinear}\includegraphics[width=50mm]{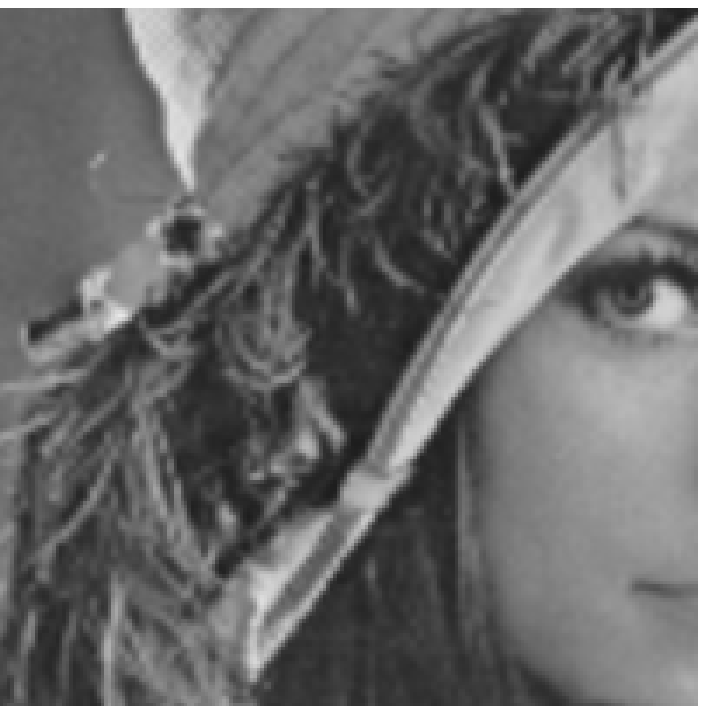}}
\subfigure[]
{\label{fig:Bicubic}\includegraphics[width=50mm]{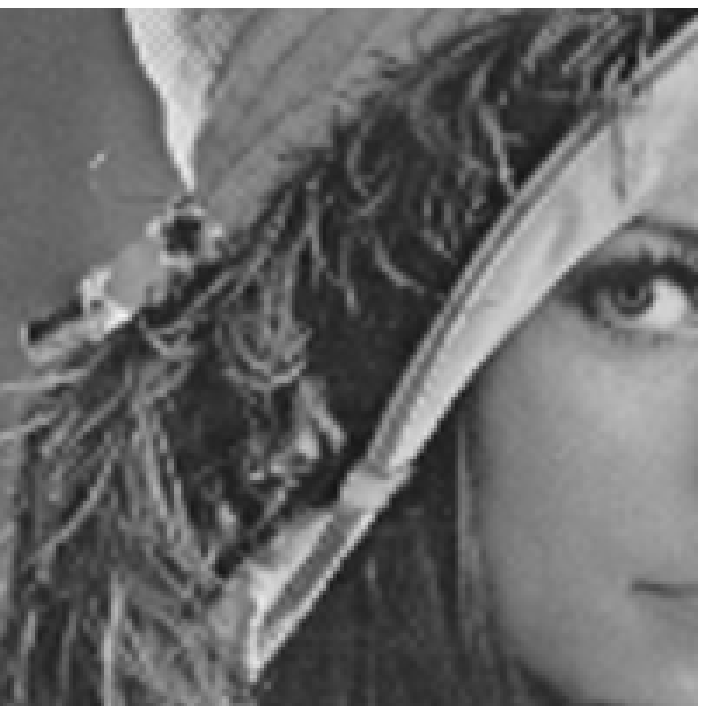}}
\subfigure[]
{\label{fig:WZPCS}\includegraphics[width=50mm]{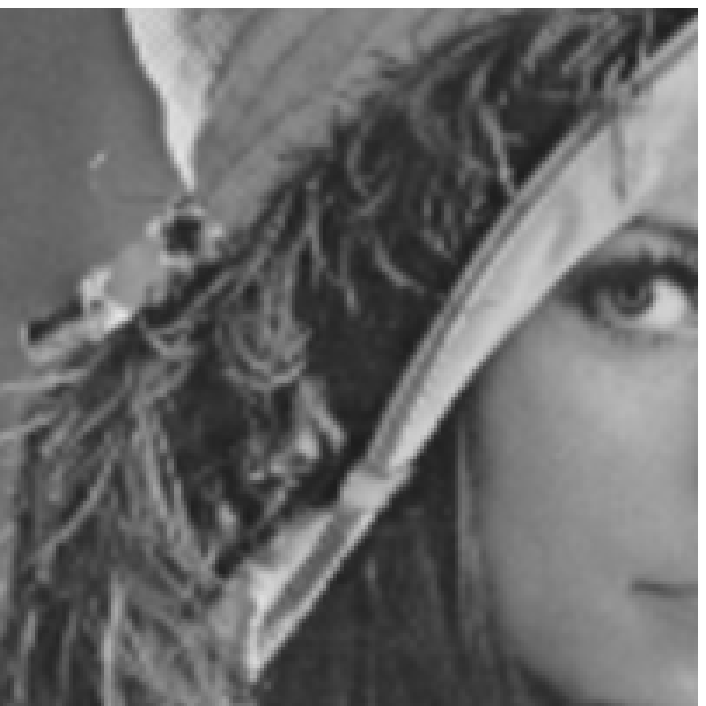}}
\subfigure[]
{\label{fig:SAI}\includegraphics[width=50mm]{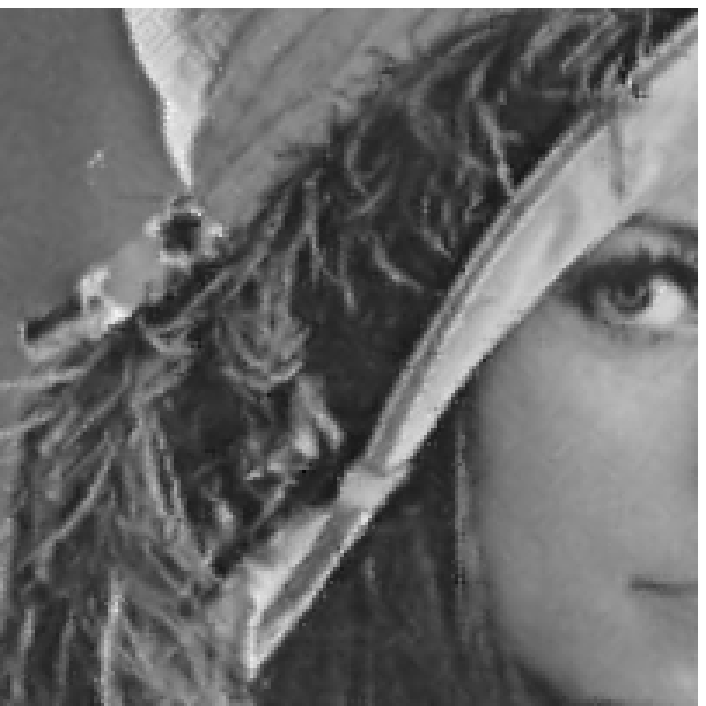}} 
\subfigure[]
{\label{fig:Opt}\includegraphics[width=50mm]{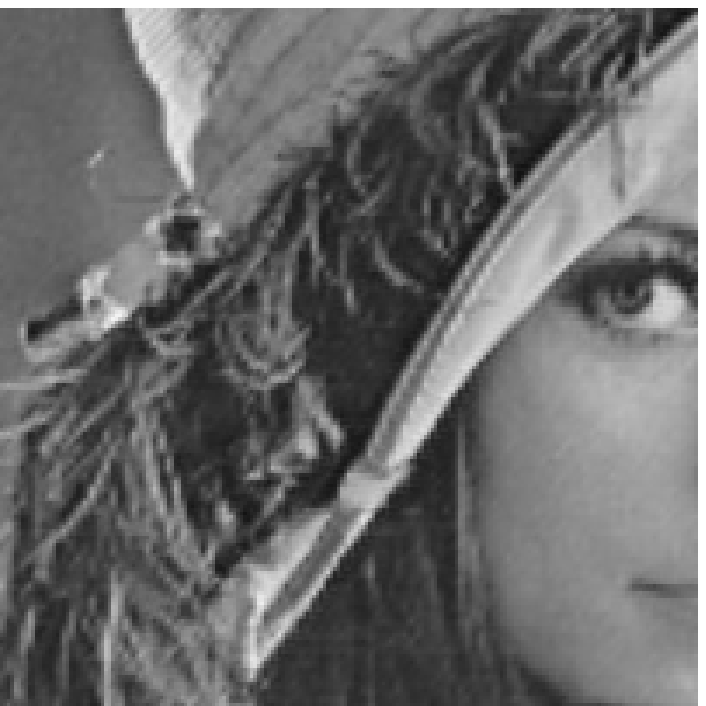}}
\caption{Comparison of different methods for the Lena image: (a) The original image, (b) bilinear interpolation, (c) bicubic Interpolation. (d) WZP Cycle-Spinning \cite{temizel2005wavelet}, (e) SAI \cite{zhang2008image}, and (f) the proposed method.
}
\vspace{-.1in}
\label{fig:Lena}
\end{figure*}

Although the PSNR value is a good measure of global quality of an image, it does not reflect the local properties. In order to present a qualitative view of various interpolation methods, we have plotted the enlarged images for a segment of the Lena test image in Fig. \ref{fig:Lena}. To highlight the differences, one could compare the texture on the top and the sharpness on the bottom edge of the hat.
%The PSNR criterion is a measure to show the global quality of an image. To  while the local inf, the spatial locations where the proposed algorithm produces significantly smaller interpolation errors than the other competing methods are plotted in Fig. \ref{fig:figure1}. The differences are more noticeable around the edge of the hat. The result of the present study compare favorably both subjectively and objectively. In addition, a wavelet scheme based on cycle-spinning interpolation has been included to provide a comparison with a powerful method operating in the wavelet domain.  

%\begin{table}[t]
%\begin{center}\label{Table4}
%\caption{SNR (dB) Results for Reconstruction of an Audio File by Various Methods (2X)}
%\begin{tabular}{|c|c|c|c|c|c|c|c|}
%\hline
%\textbf{Prefilter} & No-Filter & Anti-Aliasing & Anti-Aliasing\\\hline
%\textbf{Compared to} & Original & Original & Anti-Aliased\\\hline\hline
%\textbf{Optimized Spline% for Signal ${\rho}^3 [x,\rho^3_d]$ 
%} & 15.283 & 25.412 & 32.409\\\hline
%\textbf{Optimized Spline% for the Ideal LowPass ${\rho}^3 [sinc(t),\rho^3_d]$ 
%} & 15.176 & 26.262 & 28.212\\\hline
%\textbf{Bicubic} & 15.316 & 27.567 & 30.685\\\hline
%\end{tabular}
%\end{center}
%\vspace{-.2in}
%\end{table}

\section{Conclusion}\label{conclusion}
The interpolation problem using uniform knots is a well studied subject. In this paper, we considered the problem of optimizing the interpolation kernel for a given class of signals (represented by a filter). Although functional optimization in the continuous domain is often very difficult, we have demonstrated the equivalency of this problem with a finite dimensional linear problem which can be easily solved using linear algebra. As a special case, we considerd the class of lowpass signals which is associated with the sinc function as the optimum interpolation kernel. For the optimum compact support interpolant, we compared our function with the conventional cubic B-Spline; the simulation results indicate $1$dB improvement in the SNR of the interpolated signal (on the average) using the introduced function, and $7$dB improvement compared to the cardinal spline itself (compared to the sinc function) (Fig. \ref{fig:InterpSplines}).

%This paper has introduced a method for optimizing a compact support interpolating spline for approximating a given filter in the least square sense. In particular,it demonstrated a newly proposed method for approximating the ideal lowpass filter. The interpolation results obtained by this method are better than those obtained by the conventional solutions, such as B-splines. Simulation results show about $1$dB improvement in most of the cases. In the future, we plan to focus on the application of these optimized splines for non-uniform sampling for 1-D and 2-D signals.

\section*{Acknowledgment}
The authors would like to thank Prof. M.~Unser from EPFL and Dr.~R.~Razvan from the mathematical sciences department of Sharif university for their helpful comments.
\nocite{*}
\bibliographystyle{IEEEtran}
\bibliography{Spline.bib}

\begin{thebibliography}{10}
\providecommand{\url}[1]{#1}
\csname url@samestyle\endcsname
\providecommand{\newblock}{\relax}
\providecommand{\bibinfo}[2]{#2}
\providecommand{\BIBentrySTDinterwordspacing}{\spaceskip=0pt\relax}
\providecommand{\BIBentryALTinterwordstretchfactor}{4}
\providecommand{\BIBentryALTinterwordspacing}{\spaceskip=\fontdimen2\font plus
\BIBentryALTinterwordstretchfactor\fontdimen3\font minus
  \fontdimen4\font\relax}
\providecommand{\BIBforeignlanguage}[2]{{%
\expandafter\ifx\csname l@#1\endcsname\relax
\typeout{** WARNING: IEEEtran.bst: No hyphenation pattern has been}%
\typeout{** loaded for the language `#1'. Using the pattern for}%
\typeout{** the default language instead.}%
\else
\language=\csname l@#1\endcsname
\fi
#2}}
\providecommand{\BIBdecl}{\relax}
\BIBdecl

\bibitem{slepian1976}
D.~Slepian, ``On bandwidth,'' \emph{Proceedings of the IEEE}, vol.~64, no.~3,
  pp. 292 -- 300, 1976.

\bibitem{daubechies1988orthonormal}
I.~Daubechies, ``{Orthonormal bases of compactly supported wavelets},''
  \emph{Communications on pure and applied mathematics}, vol.~41, no.~7, pp.
  909--996, 1988.

\bibitem{strang1996wavelets}
G.~Strang and T.~Nguyen, \emph{{Wavelets and filter banks}}.\hskip 1em plus
  0.5em minus 0.4em\relax Wellesley Cambridge Pr, 1996.

\bibitem{mallat1999wavelet}
S.~Mallat, \emph{{A wavelet tour of signal processing}}.\hskip 1em plus 0.5em
  minus 0.4em\relax Academic Pr, 1999.

\bibitem{Ville2005Mul}
D.~Van De~Ville, T.~Blu, and M.~Unser, ``Isotropic polyharmonic b-splines:
  scaling functions and wavelets,'' \emph{Image Processing, IEEE Transactions
  on}, vol.~14, no.~11, pp. 1798 --1813, 2005.

\bibitem{Mallat1989Mul}
S.~Mallat, ``A theory for multiresolution signal decomposition: the wavelet
  representation,'' \emph{Pattern Analysis and Machine Intelligence, IEEE
  Transactions on}, vol.~11, no.~7, pp. 674 --693, Jul. 1989.

\bibitem{Flandrin1992Self}
P.~Flandrin, ``Wavelet analysis and synthesis of fractional brownian motion,''
  \emph{Information Theory, IEEE Transactions on}, vol.~38, no.~2, pp. 910
  --917, Mar. 1992.

\bibitem{Unser2007Self}
M.~Unser and T.~Blu, ``Self-similarity: Part i mdash;splines and operators,''
  \emph{Signal Processing, IEEE Transactions on}, vol.~55, no.~4, pp. 1352
  --1363, 2007.

\bibitem{Mallat1992Singularity}
S.~Mallat and W.~Hwang, ``Singularity detection and processing with wavelets,''
  \emph{Information Theory, IEEE Transactions on}, vol.~38, no.~2, pp. 617
  --643, Mar. 1992.

\bibitem{unser1999splines}
M.~Unser, ``{Splines: A perfect fit for signal and image processing},''
  \emph{IEEE Signal processing magazine}, vol.~16, no.~6, pp. 22--38, 1999.

\bibitem{unser1993b}
M.~Unser, A.~Aldroubi, and M.~Eden, ``{B-spline signal processing. I.
  Theory},'' \emph{IEEE transactions on signal processing}, vol.~41, no.~2, pp.
  821--833, 1993.

\bibitem{unser2000sampling}
M.~Unser, ``{Sampling-50 years after Shannon},'' \emph{Proceedings of the
  IEEE}, vol.~88, no.~4, pp. 569--587, 2000.

\bibitem{keys1981cubic}
R.~Keys, ``{Cubic convolution interpolation for digital image processing},''
  \emph{IEEE Transactions on Acoustics, Speech and Signal Processing}, vol.~29,
  no.~6, pp. 1153--1160, 1981.

\bibitem{temizel2005wavelet}
A.~Temizel, T.~Vlachos, and W.~Visioprime, ``{Wavelet domain image resolution
  enhancement using cycle-spinning},'' \emph{Electronics Letters}, vol.~41,
  no.~3, pp. 119--121, 2005.

\bibitem{zhang2008image}
X.~Zhang and X.~Wu, ``{Image interpolation by adaptive 2-D autoregressive
  modeling and soft-decision estimation.}'' \emph{IEEE transactions on image
  processing: a publication of the IEEE Signal Processing Society}, vol.~17,
  no.~6, p. 887, 2008.

\bibitem{horbelt2000spline}
S.~Horbelt, A.~Munoz, T.~Blu, and M.~Unser, ``{Spline kernels for
  continuous-space image processing},'' in \emph{IEEE INTERNATIONAL CONFERENCE
  ON ACOUSTICS SPEECH AND SIGNAL PROCESSING}, vol.~4.\hskip 1em plus 0.5em
  minus 0.4em\relax Citeseer, 2000.

\bibitem{unser13fast}
M.~Unser, A.~Aldroubi, and M.~Eden, ``{Fast B-spline Transforms for Continuous
  Image Representation and Interpolation (PDF)},'' \emph{IEEE Transactions on
  Pattern Analysis and Machine Intelligence}, vol.~13, no.~3.

\bibitem{hou1978cubic}
H.~Hou and H.~Andrews, ``{Cubic splines for image interpolation and digital
  filtering},'' \emph{IEEE Transactions on Acoustics Speech and Signal
  Processing}, vol.~26, pp. 508--517, 1978.

\bibitem{aldroubi1992cardinal}
A.~Aldroubi, M.~Unser, and M.~Eden, ``{Cardinal spline filters: Stability and
  convergence to the ideal sinc interpolator},'' \emph{Signal Processing},
  vol.~28, no.~2, pp. 127--138, 1992.

\bibitem{th^?venaz2000interpolation}
P.~Th{\'e}venaz, T.~Blu, and M.~Unser, ``{Interpolation revisited},''
  \emph{IEEE Trans. Med. Imaging}, vol.~19, no.~7, pp. 739--758, 2000.

\end{thebibliography}

\end{document}